\documentclass{aastex631}

\usepackage{threeparttable}
\usepackage{longtable}

\newcommand{\ha}{\hbox{H$\alpha$}}
\newcommand{\hb}{\hbox{H$\beta$}}

\newcommand{\sii}{\hbox{[S\,{\sc ii}]}}
\newcommand{\nii}{\hbox{[N\,{\sc ii}]}}
\newcommand{\hi}{\hbox{H\,{\sc i}}}

\begin{document}

\title{Properties of Galaxies with Counter-rotating Stellar Disks in the MaNGA Survey}

\author[0009-0005-9342-9125]{Min Bao}
\author{Zhenyu Tang}\thanks{Min Bao and Zhenyu Tang share first authorship.}
\author[0000-0003-3226-031X]{Yanmei Chen}\thanks{E-mail: \url{chenym@nju.edu.cn}.}
\affiliation{School of Astronomy and Space Science, Nanjing University, Nanjing 210023, China}
\affiliation{Key Laboratory of Modern Astronomy and Astrophysics (Nanjing University), Ministry of Education, Nanjing 210023, China}

\author[0000-0002-8614-6275]{Yong Shi}
\affiliation{Department of Astronomy, Westlake University, Hangzhou 310030, Zhejiang Province, China}

\author[0000-0002-3890-3729]{Qiusheng Gu}
\affiliation{School of Astronomy and Space Science, Nanjing University, Nanjing 210023, China}
\affiliation{Key Laboratory of Modern Astronomy and Astrophysics (Nanjing University), Ministry of Education, Nanjing 210023, China}

\begin{abstract}

Gas accretion process can fuel both star formation and black hole activity, playing a critical role in galaxy evolution. The counter-rotating structures are believed to originate from gas accretion, serving as an ideal laboratory for studying its impact on galaxy evolution. Based on the Mapping Nearby Galaxies at Apache Point Observatory (MaNGA) survey, we built a sample of 147 galaxies with counter-rotating stellar disks (CRDs). This is the largest CRD sample to date, accounting for $\sim$1.5\% of the MaNGA survey. For a subset of 138 CRDs, global stellar mass ($M_\ast$) and star formation rate (SFR) were measured in reference. We constructed a control sample with similar $M_\ast$ and SFR but lacking counter-rotating structures. The CRDs relatively exhibit more bulge-dominated morphology, lower molecular gas mass fraction and reside in less dense environment, supporting the hypothesis that they primarily originate from gas accretion. We classified 96 out of 138 CRDs into four types based on their stellar and gas kinematics following the criteria from \cite{2022ApJ...926L..13B}. There are two additional CRD types: 8 CRDs show misalignment between both stellar disks and gas disk, indicating multiple gas accretion events with differing angular momentum directions; 34 CRDs lack ionized gas emission, showing the highest $M_\ast$ among all the CRD types, which may represent a final stage of CRD evolution. We compared the radial gradients of gas-phase metallicity and stellar population properties between CRD types, and found that the impact of gas accretion on galaxy evolution primarily depends on the abundance of pre-existing gas in progenitors.

\end{abstract}


\section{Introduction} \label{sec:intro}

In the $\Lambda$CDM cosmology, the first structure formed in the Universe is the dark matter halo \citep{1984Natur.311..517B}. Subsequently, baryons fall into the potential well of dark matter halo, assembling galaxy \citep{1980MNRAS.193..189F}. Gas accretion \citep{2009MNRAS.397..208C, 2010MNRAS.409..491K, 2011MNRAS.417..666T} and galaxy merger \citep{2007ApJ...667..859D} are the primary mechanisms for galaxy assembly. The former involves the smooth accretion of external gas, while the latter refers to the merging with other galaxies. \cite{2011MNRAS.413.1373W} investigated the assembly history of Milky Way-mass dark matter haloes using Aquarius simulation series, confirming that the majority of baryons from which galaxies form are acquired through external accretion rather than mergers. \cite{2012A&A...544A..68L} employed a multi-zoom simulation to study the formation of 530 galaxies, and found that the average fraction of mass assembled by gas accretion is $\sim$77\%, indicating that gas accretion dominates galaxy assembly.

Indirect evidences of gas accretion have been identified in {\hi} observations of galaxies and their environments \citep{2008A&ARv..15..189S}, such as: (1) a large number of galaxies being surrounded by {\hi}-rich dwarfs or exhibiting {\hi} tails; (2) significant amounts of extra-planar {\hi} around nearby spiral galaxies; (3) extended and warped {\hi} layers presenting in spiral galaxies; and (4) most {\hi} disks showing morphological or dynamical asymmetries. However, gas accretion is not the only possible explanation for these features. Early observational studies aiming to directly confirm gas accretion were constrained by the sensitivity limitations of old radio telescopes \citep{2009A&A...505..559M, 2014A&A...571A..67M}. With the development of radio telescopes, it is now possible to study gas accretion processes through {\hi} emission lines in individual galaxies. For instance, \cite{2025MNRAS.540.2396S}, using {\hi} observations with high spatial resolution from MeerKAT, found that the {\hi} in NGC~5643 extends beyond the stellar disk with morphological and kinematic asymmetries, and identified an extended 30~kpc tail with counter-rotating velocities on the northern side of the disk, providing direct evidence for {\hi} gas accretion in this galaxy. Nevertheless, to date, such investigations of gas accretion processes remain limited to case studies of individual sources.

Kinematically misaligned galaxies, including gas-gas, gas-stellar or stellar-stellar components having different angular momentum (AM) directions, are widely believed to be products of gas accretion \citep{2014ASPC..486...51C}, serving as an ideal laboratory for studying its impact on galaxy evolution. The external gas can be accreted from gas-rich satellites \citep{2025arXiv250402925G} or the cosmic web \citep{2014A&ARv..22...71S}. Due to the large collision cross-section of gas components, fewer than ten galaxies with gas-gas misalignment have been observed to date, with the majority detected by long-slit spectroscopy \citep{1994ApJ...420..558B, 1994AJ....107..160F, 1996A&A...307..391P, 1999AJ....118.2184H, 2002A&A...382..488C, 2022NatAs...6.1464C}. Thanks to the development of large integral-field spectroscopy surveys, like ATLAS$^{\rm 3D}$ \citep{2011MNRAS.413..813C}, CALIFA \citep{2012A&A...538A...8S} and SAMI \citep{2012MNRAS.421..872C}, the phenomenon of gas-star misalignment was found to be ubiquitous in quiescent galaxies \citep{2011MNRAS.417..882D, 2015A&A...582A..21B, 2022MNRAS.517.2677R}. Meanwhile, samples of gas-star misalignment were also reported in star-forming galaxies \citep{2016NatCo...713269C, 2016MNRAS.463..913J, 2019MNRAS.483..458B}.

Galaxies with two counter-rotating stellar disks usually show two off-center but symmetric peaks along the major axis in the stellar velocity dispersion field, which is referred as 2$\sigma$ feature in the literature. The co-exisiting blueshifted and redshifted absorption components at the interface of the two stellar disks contribute to the broadening of absorption lines. Several galaxy cases hosting counter-rotating stellar disks (CRDs) have been studied over the past three decades \citep{1992ApJ...394L...9R, 2011MNRAS.412L.113C, 2013A&A...549A...3C, 2014A&A...570A..79P, 2016MNRAS.461.2068K, 2017MNRAS.464.4789M, 2018A&A...616A..22P}. In all these cases, the stellar disk which co-rotates with gas component is younger and has lower metallicity. Such trends support a formation scenario that the progenitor of CRDs accretes counter-rotating gas, fueling the formation of a new stellar disk which counter-rotates with the pre-existing one.

In statistic, \cite{2022MNRAS.511..139B} studied 64 galaxies with CRDs selected from $\sim$4000 galaxies in the early data release of MaNGA survey, finding the incident rate for CRDs in MaNGA is $<$5\% for ellipticals, $<$3\% for lenticulars, and $<$1\% for spirals. Through analysing the stellar populations of two stellar disks, they found the gas co-rotates with the younger disk in most cases, suggesting that CRDs form primarily via gas accretion in retrograde rotation with respect to the pre-existing stellar disk. \cite{2022ApJ...926L..13B} studied 101 galaxies with CRDs from $\sim$9500 MaNGA galaxies, classified these 101 galaxies into four types based on the features of their stellar and gas velocity fields. Under the assumption of the gas accretion formation mechanism, they propose that these distinct kinematic features are primarily dominated by two key factors, including (1) the abundance of pre-existing gas in the progenitor and (2) the efficiency in angular momentum consumption.

In this study, we build a sample of 147 CRDs from the final data release of MaNGA survey, which is the largest CRD sample to date. In Section \ref{sec:data}, we introduce the MaNGA survey and sample selection. Based on their gas and stellar kinematics, we classify the CRD sample into six types. In Section \ref{sec:results}, we analyze the global and spatially resolved properties of different types of CRDs, and make comparisons with their control samples. In Section \ref{sec:discussion}, we discuss the corresponding formation scenarios for different types of CRDs. Finally, we summarize the conclusion in Section \ref{sec:conclusion}.

\section{Data} \label{sec:data}

\subsection{MaNGA survey}

As one of the three core programs in the fourth-generation Sloan Digital Sky Survey (SDSS-IV; \citealt{2017AJ....154...28B}), MaNGA observed 10,010 unique galaxies with redshift coverage of $z\sim[0.01,~0.15]$ and flat stellar mass distribution in a range of $\log(M_{\ast}/\mathrm{M}_\odot)\sim[9,~11]$ \citep{2017AJ....154...86W, 2022ApJS..259...35A}. Using the Baryon Oscillation Spectroscopic Survey (BOSS) spectrographs \citep{2013AJ....146...32S} on the 2.5-meter telescope at the Apache Point Observatory \citep{2006AJ....131.2332G}, MaNGA employed dithered observations using integral field spectroscopy (IFS; \citealt{2015AJ....149...77D}), with diameter ranging from $12^{\prime\prime}$ (19 fibers) to $32^{\prime\prime}$ (127 fibers). Two dual-channel BOSS spectrographs provided wide wavelength coverage of $\lambda\sim[3600,~10,300]$~\AA,~with a median spectral resolution of $R\sim2000$.

The data products in this study are obtained from the final data release of MaNGA survey. For each MaNGA galaxy, the global stellar mass and star formation rate are obtained from \cite{2015ApJS..219....8C}, which are estimated through spectral energy distribution (SED) fitting combining SDSS and WISE photometry. We refer to Figures~3 and 5 of \cite{2015ApJS..219....8C} for example SED fitting results of massive, quiescent galaxies and blue, star-forming galaxies. The bulge-to-total ratio (B/T) is gathered from the MaNGA PyMorph DR17 photometric catalog \citep{2019MNRAS.483.2057F}, derived through S$\acute{\rm e}$rsic + Exponential fitting to the 2D surface brightness profile of SDSS image using \texttt{PYMORPH}. Other global properties, including photometric position angle ($\phi$), effective radius ($Re$) and axial ratio ($b/a$) are extracted from NASA-Sloan Atlas catalog (NSA, \citealt{2011AJ....142...31B}), based on one-component, two-dimensional S$\acute{\rm e}$rsic fitting. The spatially resolved light-weighted stellar metallicity ([Z/H]) is provided by the \texttt{pyPipe3D} pipeline \citep{2022ApJS..262...36S}, which serves as a proxy for the metal enrichment cycle. This quantity is derived by fitting the stellar continuum with simple stellar population (SSP) models. Other spatially resolved properties, including emission line flux, gas (or stellar) velocity and velocity dispersion, 4000~\AA~break (D$_n$4000) defined as the flux ratio between two narrow bands of 3850-3950~\AA~and 4000-4100~\AA, are obtained from MaNGA data analysis pipeline (DAP, \citealt{2019AJ....158..231W}).

DAP is a survey-led software package for analysing the spatially resolved MaNGA spectra to generate high-level data products. It performs two full-spectrum fittings for each spectrum with $g$-band signal-to-noise ratio ($\rm S/N$) higher than 1, both using penalized pixel-fitting (pPXF; \citealt{2017MNRAS.466..798C}), but with different templates. The templates used to determine the stellar kinematics are from MILESHC \citep{2006MNRAS.371..703S}, and the templates used to fit the stellar continuum during the emission-line fitting module are from MASTARSSP \citep{2019ApJ...883..175Y}. MILESHC/MASTARSSP is a stellar-template library constructed by hierarchically-clustering the MILES/MaStar stellar library. In the first full-spectrum fitting, the stellar continuum is modeled as a linear combination of MILESHC templates, convolved with the line-of-sight velocity distribution (LOSVD; \citealt{2004PASP..116..138C, 2017MNRAS.464.4789M}), to determine the stellar kinematics. The first two moments of the LOSVD correspond to velocity and velocity dispersion of stellar component. In the second full-spectrum fitting, all the emission lines are simultaneously modeled by Gaussian profiles using pPXF. Each emission line can be modeled as a Gaussian profile characterized by its flux, line center and width, with all the emission line centers tied together in the velocity space. The velocity and velocity dispersion of gas component are then derived from the modeled line center and width.

\subsection{The CRD sample}

\begin{figure*}[h]
    \centering\resizebox{1\textwidth}{!}{\includegraphics{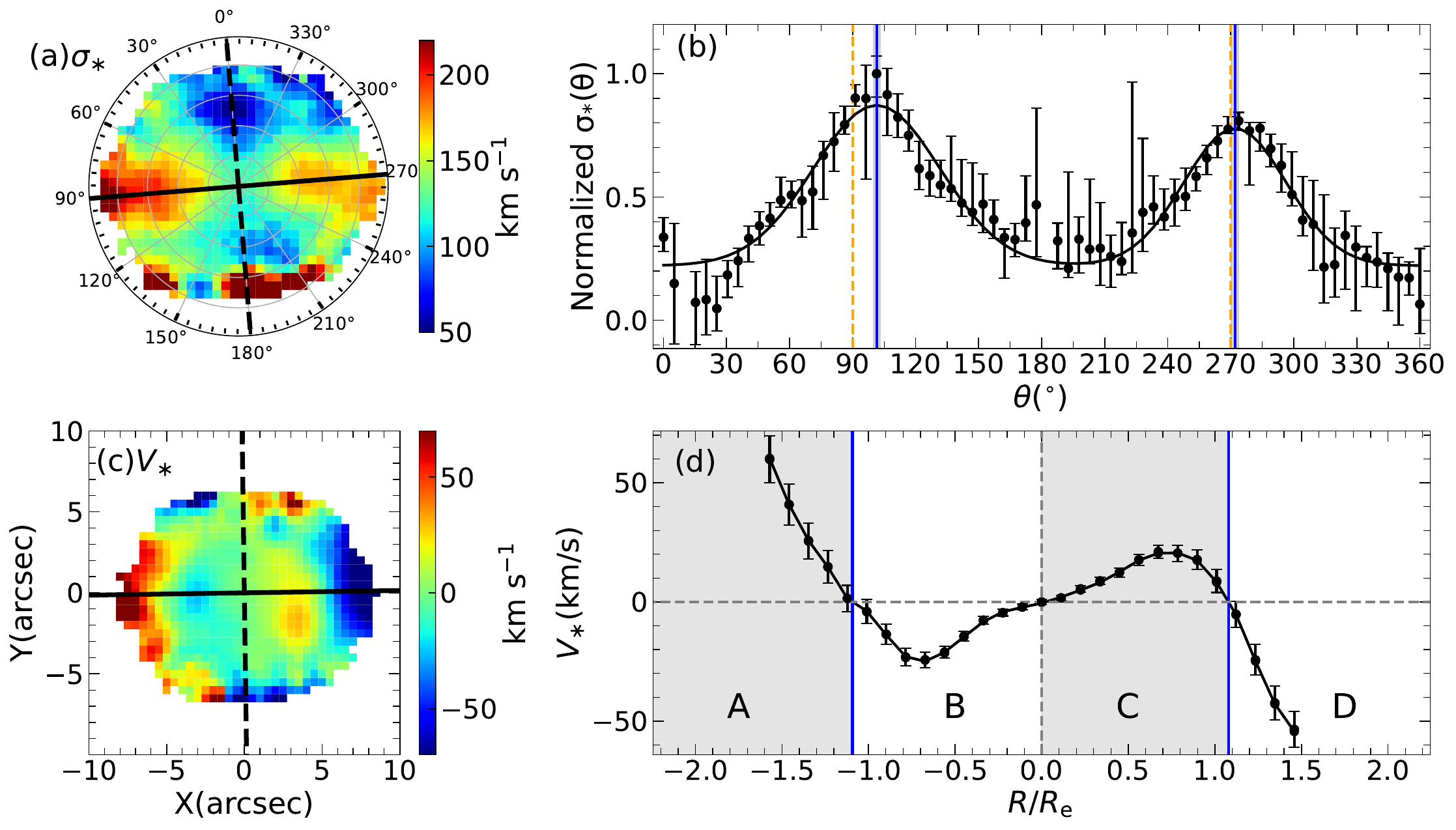}}
   \caption{Two examples of our sample selection. (a) The stellar velocity dispersion field in polar coordinate. The black solid and dashed lines show the photometric major and minor axes. (b) Normalized $\sigma_{\ast}$ as a function of $\theta$. The black circle represents the median $\sigma_{\ast}$ in each sector, with the black vertical bar showing the $\pm 1 \sigma$ scattering range. The black curve shows the best-fitting double Gaussian model. The vertical blue lines and grey-shaded areas mark the directions of two peaks and corresponding errors. The vertical orange dashed lines mark $\theta = 90^{\circ}$ and $\theta = 270^{\circ}$, the photometric major axis. (c) The stellar velocity field. The black solid and dashed lines show the photometric major and minor axes. (d) $v_\ast$ along major axis. The black circles show the $v_\ast$ as a function of radius, with the black vertical bar showing the $v_\ast$ error. The black curve shows the best result of polynomial fitting. The horizontal grey dashed line marks $v_\ast = 0~\rm km~s^{-1}$. The vertical gray dashed line marks the radius of central velocity-zero point. The vertical blue solid lines mark the radii of the other two velocity-zero points.}
   \label{fig:method}
\end{figure*}

CRDs exhibit one or both of the following characteristics: (1) the presence of 2$\sigma$ feature in the stellar velocity dispersion ($\sigma_{\ast}$) field; (2) the clear counter-rotation (with position angle oﬀset higher than $150^{\circ}$) between inner and outer stellar disks in the stellar velocity ($v_\ast$) field. The upper panels of Figure~\ref{fig:method} display an example of CRDs with 2$\sigma$ feature. Figure~\ref{fig:method}(a) displays the $\sigma_{\ast}$ field in polar coordinate, with the black solid and dashed lines representing the photometric major and minor axes. The upper half of minor axis is set as $\theta = 0^{\circ}$, with increasing value of $\theta$ in the counter-clockwise direction. To analyze the $\sigma_{\ast}$ distribution, we divide the $\sigma_{\ast}$ field into circular sectors with a sector width of $\Delta\theta = 5^{\circ}$. Figure~\ref{fig:method}(b) shows the $\sigma_{\ast}$ as a function of $\theta$, with the peak value normalized to 1. The black circle represents the median $\sigma_{\ast}$ in each sector, and the error bar shows the $\pm 1 \sigma$ scattering range. To search for the $\sigma_{\ast}$ enhanced region, we fit these black circles with a double-Gaussian model using the Python-based package \texttt{CURVE\_FIT}. The black profile in Figure~\ref{fig:method}(b) is the best-fitting Gaussian model. The two peaks locate at $\theta_1 = 101.5 \pm 1.6 ^{\circ}$ and $\theta_2 = 272.0 \pm 1.7 ^{\circ}$, marked by the vertical blue lines. We select 1267 candidates satisfying $|\theta_1 - 90^{\circ}| \leq \theta_{c}$ and $|\theta_2 - 270^{\circ}| \leq \theta_{c}$, where $\theta_{c} = 50^{\circ}$. This criterion ensures that the 2$\sigma$ feature is located around the major axis. Through visual inspection, we exclude candidates with galaxy pair within the same MaNGA field-of-view, ongoing merger, galaxies with irregular morphology, or non-smooth $\sigma_\ast$ pattern, leaving 135 CRDs exhibiting 2$\sigma$ feature.

The lower panels of Figure~\ref{fig:method} display an example of CRDs with clear counter-rotation between inner and outer stellar disks. Figure~\ref{fig:method}(c) displays the $v_\ast$ field, with the black solid and dashed lines representing the photometric major and minor axes. Figure~\ref{fig:method}(d) displays the $v_\ast$ along major axis, with the black circles showing $v_\ast$ as a function of radius. We perform polynomial fitting for these black circles, with the best-fitting result shown as a black curve. On this curve, we find three positions with $v_\ast = 0$, one at the galaxy center, one on the positive simi-major axis, and another on the negative simi-major axis. These three zero-velocity points (marked by vertical lines) divide the curve into four segments (A/B/C/D), and the directions of velocity in two adjacent segments are opposite. We select 448 candidates satisfying these criteria. We visually inspect these galaxies to delete galaxy pairs within the field-of-view, ongoing mergers, irregular morphology, or disturbed $v_\ast$ pattern, leaving 45 CRDs exhibiting counter-rotation between inner and outer stellar disks. Thirty-three out of these 45 CRDs show both counter-rotation and 2$\sigma$ features. Thus our final sample include 147 CRDs, which accounts for 1.5\% of the MaNGA survey. The basic physical properties of these galaxies are provided in the table in appendix, and the corresponding kinematic maps can be accessed on Marvin Web\footnote{https://magrathea.sdss.org/marvin/} using the MaNGA-ID or Plate-IFU. We cross-match these galaxies with the catalog provided by \cite{2015ApJS..219....8C} to obtain their global stellar mass and star formation rate. These properties were derived from SED fitting using a combination of SDSS and WISE photometry. This process leaves a subset of 138 CRDs for our subsequent analysis.

\begin{figure*}[h]
     \centering\resizebox{1\textwidth}{!}{\includegraphics{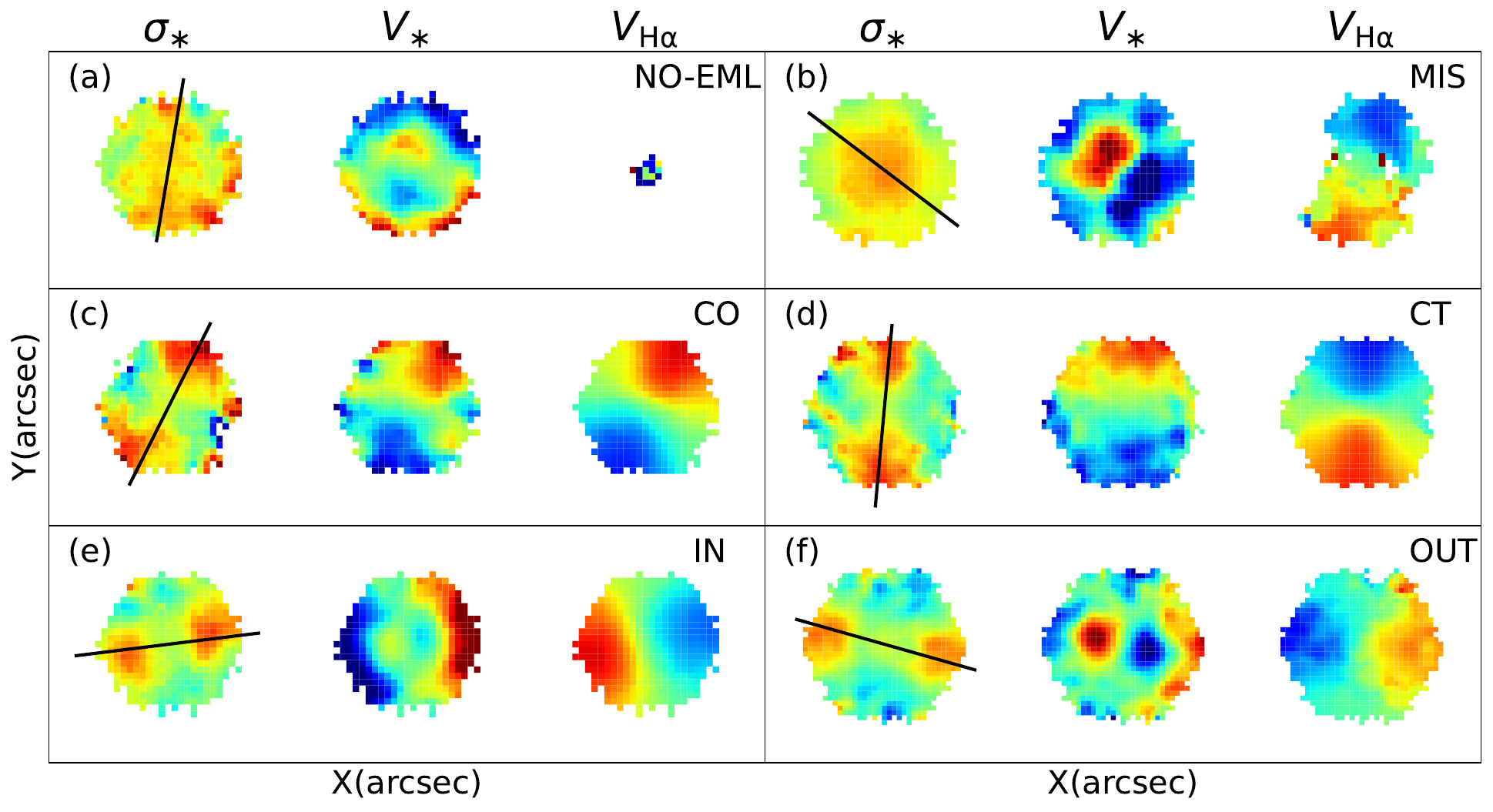}}
    \caption{Examples of different types of CRDs. (a) Type~NO-EML, where the gas emission can be neglected. (b) Type~MIS, where both stellar disks are misaligned with the gas disk. (c) Type~CO, where one stellar disk overshines the other, with the brighter stellar disk co-rotating with the gas disk. (d) Type~CT, where one stellar disk overshines the other, with the brighter stellar disk counter-rotating with the gas disk. (e) Type~IN, where the inner stellar disk is co-rotating with the gas disk. (f) Type~OUT, where the outer stellar disk is co-rotating with the gas disk. In all the panels, three subplots sequentially represent stellar velocity dispersion field, stellar velocity field, and gas velocity field traced by {\ha}. The black solid line in each stellar velocity dispersion field represents the photometric major axis.}
    \label{fig:cases}
\end{figure*}

The CRD sample is classified into six types based on their stellar and gas kinematics. Figure~\ref{fig:cases} shows six examples of corresponding types. In each panel, the three maps show stellar velocity dispersion field, stellar velocity field, and gas velocity field traced by {\ha}. Figure~\ref{fig:cases}(a) displays an example of 34 CRDs classified as Type~NO-EML, in which over 75\% spaxels within $1.5~Re$ have H$\alpha$ $\rm S/N$ lower than 3, thus we can not get robust gas velocity fields. The black solid line in the stellar velocity dispersion field represents the photometric major axis. In this study, we require median spectral $\rm S/N > 3$ for reliable stellar properties and emission-line $\rm S/N > 3$ for reliable gas properties in all the spatially resolved analyses. Figure~\ref{fig:cases}(b) displays an example of eight CRDs classified as Type~MIS, in which both stellar disks are misaligned with the gas disk. We use \texttt{KINEMETRY} package \citep{2006MNRAS.366..787K} to fit the position angle for gas disk ($\Phi_{\rm gas}$), and the position angles for inner and outer stellar disks ($\Phi_{\ast}$), respectively. Type~MIS satisfy the criterion that position angle offsets $\Delta\Phi = |\Phi_{\ast} - \Phi_{\rm gas}| > 30^{\circ}$ for both inner and outer stellar disks.

\begin{figure*}[h]
    \centering\resizebox{0.7\textwidth}{!}{\includegraphics{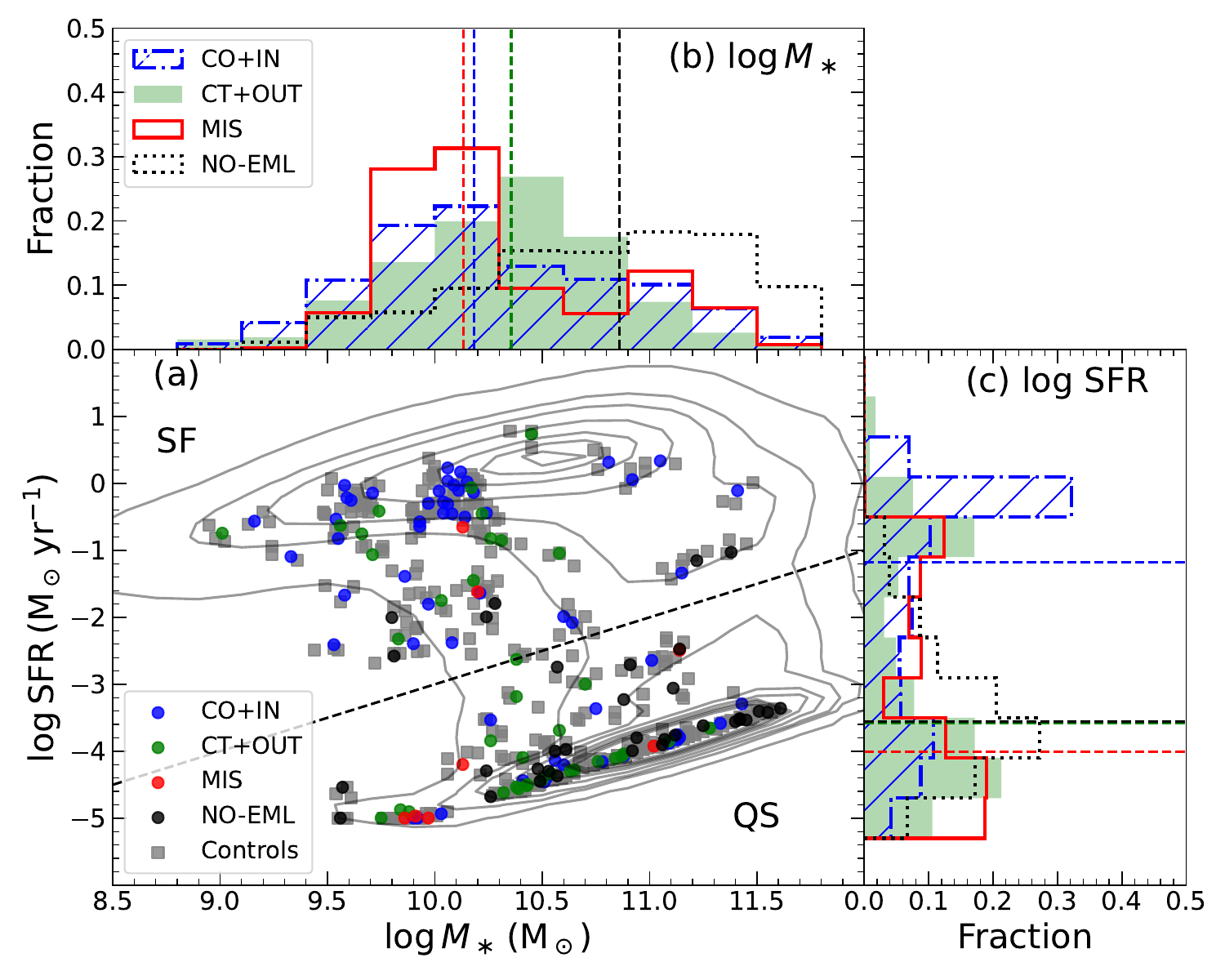}}
   \caption{Distribution of CRD sample in the $\log M_\ast$ - $\log \rm SFR$ plane. (a) The grey contour shows the distribution of SDSS spectroscopic galaxy sample from \cite{2015ApJS..219....8C} catalog. The blue, green, red and black circles show the distributions of Type~CO+IN, Type~CT+OUT, Type~MIS and Type~NO-EML, respectively. The black dashed line classifies galaxies into SF and QS sequence. (b) Distribution of stellar mass. (c) Distribution of star formation rate. In panels (b) and (c), the blue, green, red and black histograms show the $M_\ast$ and SFR distributions of Type~CO+IN, Type~CT+OUT, Type~MIS and Type~NO-EML, respectively. The blue, green, red and black vertical lines show the corresponding median values.}
   \label{fig:M_SFR}
\end{figure*}

The remaining 96 galaxies are classified into four types following the criteria from \cite{2022ApJ...926L..13B}. Sixty-six of them exhibit 2$\sigma$ features in their stellar velocity dispersion fields, but lack of evidence for two counter-rotating stellar disks in the stellar velocity fields. This phenomenon occurs because one of the two stellar disks overshines the other, and the brighter disk dominates the stellar rotation. We classify these 66 galaxies into 41 Type~CO in which the stellar and gas disks are co-rotating (with position angle oﬀset lower than $30^{\circ}$), and 25 Type~CT in which the gas and stellar disks are counter-rotating with each other. Figures~\ref{fig:cases}(c) and \ref{fig:cases}(d) show examples of Type~CO and Type~CT, respectively. The other $(96 - 66 = )~30$ galaxies clearly show two counter-rotating stellar disks in the stellar velocity fields. We separate these 30 CRDs into 17 Type~IN and 13 Type~OUT. The gas disk is co-rotating with the inner stellar disk in Type~IN (Figure~\ref{fig:cases}e), while it is co-rotating with the outer stellar disk in Type~OUT (Figure~\ref{fig:cases}f). \cite{2022ApJ...926L..13B} found that these four types have different stellar population age, with Type~CO and Type~IN being relatively younger. Type~CO and Type~IN are dominated by SF galaxies with abundant pre-existing gas. The effective interaction between pre-existing and accreted gas can trigger active star formation, resulting in younger stellar population age. Meanwhile, Type~CT and Type~OUT are primarily QS galaxies with lower amount of pre-existing gas. For the following analysis, we combine Type~CO and Type~IN together as Type~CO+IN, and Type~CT and Type~OUT as Type~CT+OUT.

To understand the formation mechanisms for CRDs, we build a control sample without any counter-rotation between gas and/or stellar components for comparison. Three control galaxies are closely matched to each CRDs in two dimensional parameter space of global stellar mass ($|\Delta \log M_{\ast}| < 0.1$) and star formation rate ($|\Delta \log \rm SFR| < 0.2$). Figure~\ref{fig:M_SFR}(a) displays the distributions of galaxies in the $\log M_\ast - \log \rm SFR$ plane. The grey contour represents the galaxies with robust global stellar mass and SFR measurements in the catalog of \cite{2015ApJS..219....8C}. The black dashed line separates these galaxies into star-forming (SF) and quiescent (QS) sequence. The blue, green, red and black circles represent Type~CO+IN, Type~CT+OUT, Type~MIS and Type~NO-EML, respectively. The grey squares represent the control sample. Table \ref{tbl:sequence} lists the numbers and fractions of SF galaxies for different types of CRDs. The first column lists the type names, followed by the total number of galaxies for each type in the second column. The third and forth columns provide the numbers of SF galaxies and their fractions in the relevant types of CRDs. It turns out that Type~CO+IN are dominated by SF galaxies, whereas Type~CT+OUT, Type~NO-EML and Type~MIS are dominated by QS galaxies. The SF fraction declines from 67.2\% in Type~CO+IN to 17.6\% in Type~NO-EML.

\begin{table*}
    \caption{The numbers and fractions of SF galaxies for different types of CRDs.}
    \centering          
    \begin{tabular}{c c c c}     
    \hline\hline       
    Type of CRDs & Total number &    Number of SF galaxies    & Fraction of SF galaxies \\ 
    \hline                    
       Type~CO+IN                                   &  58  &  39  &  67.2\%  \\
       Type~CT+OUT                                  &  38  &  14  &  36.8\%  \\
       Type~MIS                                     &   8  &   2  &  25.0\%  \\
       Type~NO-EML                                  &  34  &   6  &  17.6\%  \\
    \hline
    \label{tbl:sequence}          
    \end{tabular}
\end{table*}

Figures~\ref{fig:M_SFR}(b) and \ref{fig:M_SFR}(c) display the $\log M_\ast$ and $\log \rm SFR$ distributions for different types of CRDs, with blue, green, red and black histograms representing Type~CO+IN, Type~CT+OUT, Type~MIS and Type~NO-EML, respectively. The blue, green, red and black vertical (or horizontal) lines show the median values of $\log M_\ast$ (or $\log \rm SFR$) for corresponding distributions. In Figure~\ref{fig:M_SFR}(b), Type~NO-EML exhibit the highest stellar mass with a median value of $\log (M_\ast / \mathrm{M_\odot}) \sim 10.9$, while Type~CT+OUT are about 0.5~dex lower. Type~CO+IN and Type~MIS show comparable stellar mass, both having the lowest median values of $\log (M_\ast / \mathrm{M_\odot}) \sim 10.2$. In Figure~\ref{fig:M_SFR}(c), Type~CO+IN display the highest star formation rate with a median value of $\log (\mathrm{SFR} / \mathrm{M_\odot~yr^{-1}}) \sim -1.2$. The star formation rate of Type~CT+OUT and Type~MIS is 2.4~dex lower than that of Type~CO+IN. Although Type~MIS have stellar mass comparable to Type~CO+IN, it exhibits the lowest star formation rate of $\log (\mathrm{SFR} / \mathrm{M_\odot~yr^{-1}}) \sim -4.0$.

\section{Results} \label{sec:results}

\subsection{Global properties}

We compare the global properties between the CRD sample and their control sample, including galaxy morphology traced by bulge-to-total ratio, molecular gas content scaled from Balmer decrement, and galaxy environment indicated by neighbour number and tidal strength parameter. Under the assumption of gas accretion formation mechanism, the interaction between pre-exisiting and accreted gas can redistribute angular momentum and trigger gas inflows. We are interested in: (1) the impact of gas accretion on galaxy evolution, (2) how it modulates the molecular gas content, and (3) whether CRDs prefer to reside in certain environments.

\subsubsection{Morphology}

\begin{figure*}[h]
    \centering\resizebox{1.0\textwidth}{!}{\includegraphics{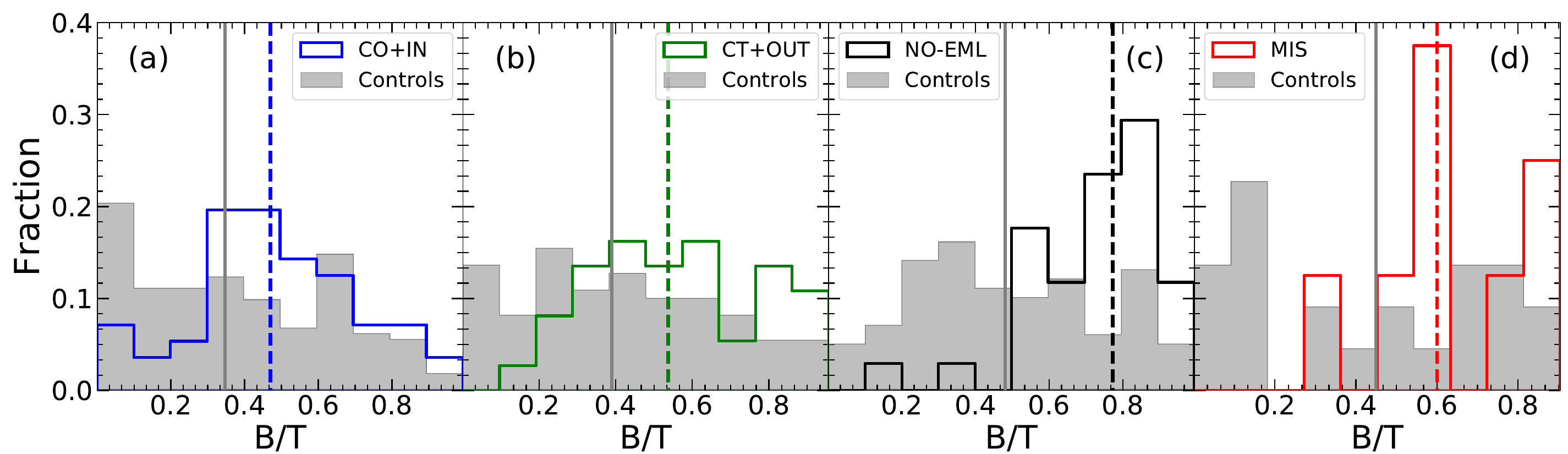}}
   \caption{The distributions of bulge-to-total ratio. Four panels sequentially display the B/T distributions for Type~CO+IN, Type~CT+OUT, Type~NO-EML, Type~MIS and their control samples. The colorful histograms show the B/T distributions for diﬀerent CRD types. The grey histograms show the B/T distributions for their control samples. The colorful and grey vertical lines show the corresponding median values.}
   \label{fig:morphology}
\end{figure*}

Figure~\ref{fig:morphology} compares the distributions of bulge-to-total ratio between Type~CO+IN (panel a), Type~CT+OUT (panel b), Type~NO-EML (panel c), Type~MIS (panel d) and their control samples. In each panel, the colored histogram represents the CRD sample, while the gray histogram represents the control sample. The vertical lines indicate the median values of corresponding distributions.

All the CRD types in Figure~\ref{fig:morphology} show larger bulge-to-total ratios than their control samples, indicating the more bulge-domimated morphology of these CRDs. In the four panels, the median B/T of the control samples are below 0.5, suggesting that the bulge tend to contribute less than half of the total light in these galaxies. However, except for Type~CO+IN whose median B/T value is slightly below 0.5 ($\sim 0.47$), the median B/T of Type~CT+OUT ($\sim 0.54$), Type~NO-EML ($\sim 0.77$), and Type~MIS ($\sim 0.60$) all exceed 0.5. These CRDs are believed to form mainly by accreting external gas. As suggested by \cite{2016NatCo...713269C}, AM redistribution occurs during interactions between pre-existing and accreted gas, driving gas inflows and central star formation. This enhanced star formation in the center then contributes to the growth of bulge, and alters the system’s light distribution.

\subsubsection{Gas content}

Current studies suggest that CRDs accrete gas from external environments, which may result in elevated gas masses. Conversely, interactions between pre-existing and accreted gas can drive AM redistribution and gas inflows, leading to rapid gas-to-star transformation. These two competing processes jointly modulate the gas content in CRDs. The molecular gas surface density ($\mathit{\Sigma}_{\rm mol}$) is estimated as $\mathit{\Sigma}_{\rm mol} \sim 23~M_\odot~\mathrm{pc}^{-2}~A_V$, where $A_V$ is $V$-band extinction calculated from the Balmer decrement \citep{2020MNRAS.492.2651B}. The Balmer decrement is defined as the flux ratio of {\ha} to {\hb} Balmer lines ({\ha}/{\hb}). The fluxes of {\ha} and {\hb} are obtained from the MaNGA DAP, which are derived through Gaussian profile fitting using pPXF. The total molecular gas mass ($M_{\rm mol}$) is obtained as $\Sigma_i (S_{\rm spaxel}^i \times \mathit{\Sigma}_{\rm mol}^i)$, where $S_{\rm spaxel}^i$ is the physical area and $i$ means the $i$-th spaxel.

\begin{figure*}[h]
    \centering\resizebox{1\textwidth}{!}{\includegraphics{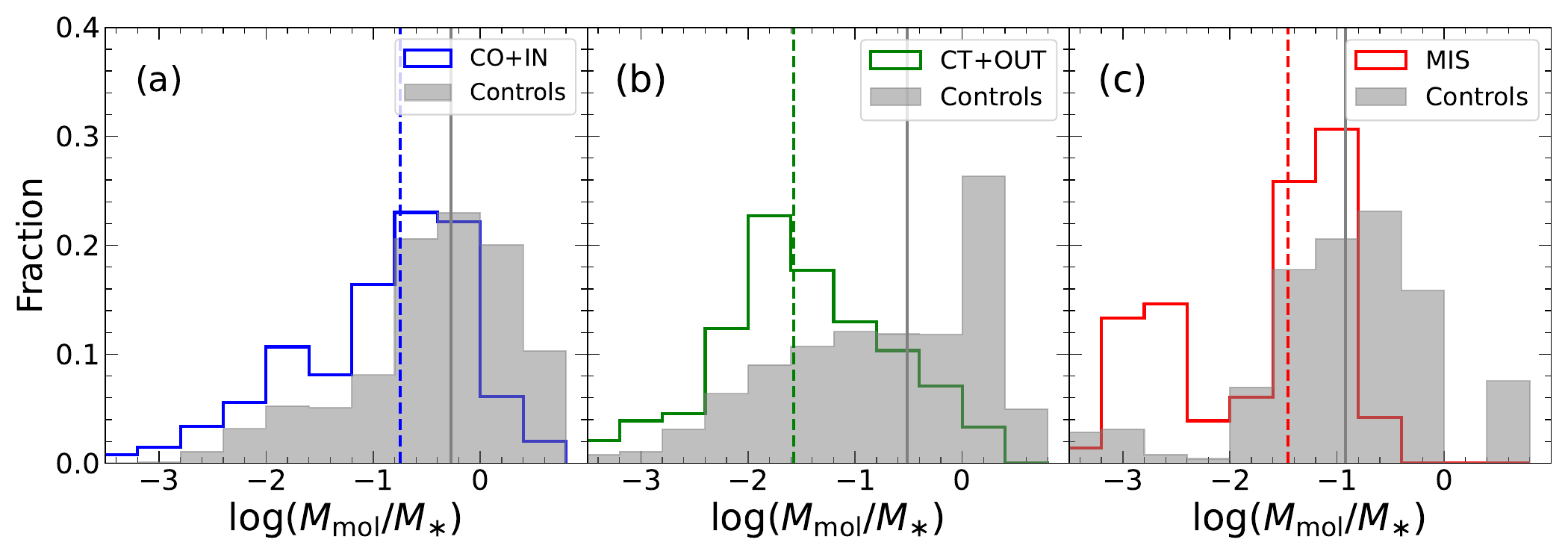}}
   \caption{The distributions of molecular gas mass fraction. Three panels sequentially display the $\log(M_{\rm mol} / M_\ast)$ distributions for Type~CO+IN, Type~CT+OUT, Type~MIS and their control samples. The colorful histograms show the $\log(M_{\rm mol} / M_\ast)$ distributions for different CRD types. The grey histograms show the $\log(M_{\rm mol} / M_\ast)$ distributions for their control samples. The colorful and grey vertical lines show the corresponding median values.}
   \label{fig:Mgas}
\end{figure*}

Figure~\ref{fig:Mgas} displays the distributions of molecular gas mass fraction defined as $\log(M_{\rm mol} / M_\ast)$ between CRDs and control samples, with colors coded in the same way as Figure~\ref{fig:morphology}. It is clearly shown in Figure~\ref{fig:Mgas} that all the three types of CRDs have lower molecular gas mass fraction than their control samples. The median value of molecular gas mass fraction of Type~CO+IN is $\sim$0.5~dex lower than their control sample in Figure~\ref{fig:Mgas}(a). The difference between Type~CT+OUT and their control sample increases to $\sim$1.1~dex in Figure~\ref{fig:Mgas}(b). And the difference between Type~MIS and their control sample is $\sim$0.6~dex, which falls between those of the other two types.

Similar results are reported in the gas-star misaligned samples selected from MaNGA survey \citep{2020MNRAS.492.1869D, 2022MNRAS.515.5081Z} as well as that from IllustrisTNG simulation \citep{2021MNRAS.503..726L}. The lower molecular gas mass fraction could naturally result from efficient consumption. Molecular gas, the raw material for star formation, is efficiently consumed by the enhanced star formation in these galaxies, which is triggered by the interaction between pre-existing and accreted gas.

\subsubsection{Large-scale environment}

In the mechanism of external gas accretion, the external gas primarily resides outside the virial radius of the dark-matter halo that hosts the galaxy, and by accretion over cosmic time it becomes part of the baryon pool that supplies the evolution of this galaxy \citep{2012A&A...538A...8S}. The origin of external gas involves sources such as the intergalactic medium (IGM; \citealt{2025ApJ...992..169C}), satellites \citep{2026ApJ...997...54L}, and cosmic web \citep{2025ApJ...982L..29B}. Under such a mechanism $-$ the formation of CRDs occurs primarily through the external gas accretion $-$ we would expect these galaxies to reside predominantly in sparse environments conducive to gas accretion. Conversely, in dense environments, mechanisms such as gas stripping or halo heating may suppress this process.

Based on the ATLAS$^{\rm 3D}$ survey, \cite{2011MNRAS.417..882D} demonstrated that early-type galaxies in dense groups and clusters are more likely to exhibit kinematic alignment between gas and stars, whereas misalignments tend to dominate in field galaxies. Focusing on 66 gas-star misaligned galaxies with diverse star formation types selected from the early data release of the MaNGA survey, \cite{2016MNRAS.463..913J} further showed that these systems all reside in relatively isolated environments. With a sample size about ten times larger, \cite{2022MNRAS.515.5081Z} revisited the environmental dependence of misaligned galaxies and corroborated this result. Here, we compare the environmental parameters (the neighbour number and the tidal strength parameter) between our CRDs and their control samples.

\begin{figure*}[h]
    \centering\resizebox{1\textwidth}{!}{\includegraphics{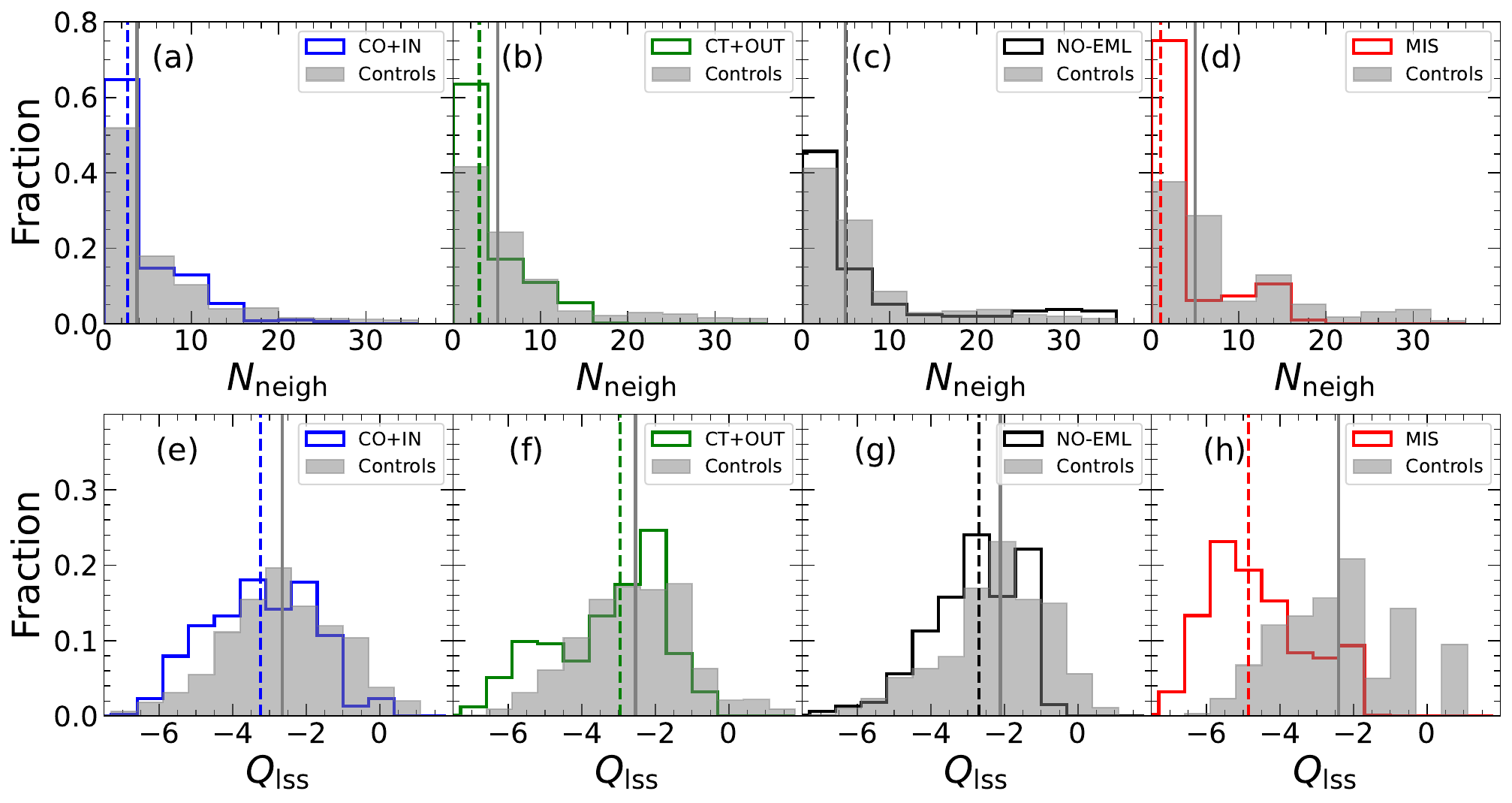}}
   \caption{The distributions of environmental parameters. The top and bottom rows display the distributions of neighbour number and tidal strength parameter, respectively. Four panels in each row sequentially display the distributions for Type~CO+IN, Type~CT+OUT, Type~NO-EML, Type~MIS and their control samples. The colorful histograms show the distributions for different CRD types. The grey histograms show the distributions for their control samples. The colorful and grey vertical lines show the corresponding median values.}
   \label{fig:Qlss}
\end{figure*}

The neighbour number ($N$neigh) is defined as the number of neighbours with $r$-band absolute magnitude brighter than $-$19.5 within 500~km~s$^{-1}$ line-of-sight velocity and 1~Mpc projected distance in a volume-limited sample with $z < 0.15$ \citep{2015AA...578A.110A}. The tidal strength parameter $Q_{\rm lss}$ is a description of the total gravitational interaction strength by all the neighbours within this limited volume. The estimation method is described in \cite{2015AA...578A.110A}, and the result is released in the Galaxy Environment for MaNGA Value Added Catalog (GEMA-VAC)\footnote{https://data.sdss.org/datamodel/files/MANGA\_GEMA/GEMA\_VER/GEMA.html}. It is defined as
\begin{equation}
Q_{\rm{lss}} = \log [\Sigma\frac{M_i}{M_t}(\frac{D_t}{d_i})^3],
\label{con:S/N}
\end{equation}
where $M_t$ and $M_i$ are the stellar mass of the primary galaxy and its $i$-th neighbor, $d_i$ is the projected distance between the primary galaxy and the $i$-th neighbor, and $D_t$ is the diameter of the primary galaxy. 

Figure~\ref{fig:Qlss} displays the distributions of $N$neigh in the top panels and $Q_{\rm lss}$ in the bottom panels, with colors coded in the same way as Figure~\ref{fig:morphology}. From left to right, the CRD types are illustrated Type~CO+IN, Type~CT+OUT, Type~NO-EML and Type~MIS. The two vertical lines in each panel mark the median values of corresponding distributions. All these types of CRDs tend to reside in more isolated environments with lower values of $Q_{\rm lss}$ than their control samples. Similar trends are also found in $N$neigh for Type~CO+IN, Type~CT+OUT and Type~MIS, but with less significance. Figures~\ref{fig:Qlss}(d) and \ref{fig:Qlss}(h) show the largest difference between Type~MIS and their control of all the CRD types. Type~MIS host two counter-rotating stellar disks, with both stellar disks rotating in directions inconsistent with the gas disk. This configuration suggests that such galaxies have experienced multiple episodes of external gas accretion. Given the complexity of their formation processes, we will conduct a detailed investigation of these galaxies in future work.

\subsection{Spatially resolved properties}

In this section, we study the spatially resolved properties of the CRD sample, including gas-phase metallicity, stellar population age traced by 4000~\AA~break and light-weighted stellar metallicity. The dilution of gas-phase metallicity can serve as evidence of CRDs accreting gas from the external environments. Meanwhile, combining the stellar population age and light-weighted stellar metallicity, we can understand the star formation and metal enrichment cycle under the influence of gas accretion.

\subsubsection{Gas-phase metallicity}

Gas-phase metallicity is one of the most fundamental physical properties of galaxies, reflecting both the amount of gas reprocessed by stellar nucleosynthesis and the exchange of gas between galaxies and surrounding environments. CRDs are widely regarded as key tracers of evolution dominated by external processes. External mechanisms $-$ such as minor/major mergers or gas accretion $-$ can bring in gas with metallicity distinct from the pre-existing gas within galaxies. By comparing the gas-phase metallicity of CRDs with that of their control samples, we can gain critical insights into the origin of the external gas and elucidate how gas accretion processes shape galaxy evolution.

Strong-line abundance diagnostics, developed based on stellar population synthesis and photoionization models, are primarily applicable to H\textsc{ii} regions \citep{2002ApJS..142...35K}. However, in most spaxels of QS galaxies, the gas is not excited by star formation. Following \cite{2016MNRAS.463..913J}, we therefore adopt the $\mathrm{N2S2} = \log({\nii}\lambda6583 / ({\sii}\lambda6717 + {\sii}\lambda6731))$ as an alternative gas-phase metallicity indicator. \cite{2016ApSS.361...61D} and \cite{2016ApJ...823L..24K} proposed that this ratio serves as a proxy for the N/O ratio and exhibits a tight correlation with O/H abundance at 12 + log(O/H) $>$ 8.0. Additionally, the two emission lines are closely spaced in wavelength, making dust extinction corrections negligible. The gas-phase metallicity is calculated following the scaling relation from \cite{2022A&A...659A.125S}: $12 + \log(\rm O/H) = 8.875 + 0.827 \times N2S2 - 0.288 \times N2S2^2$.

\begin{figure*}[h]
    \centering\resizebox{1\textwidth}{!}{\includegraphics{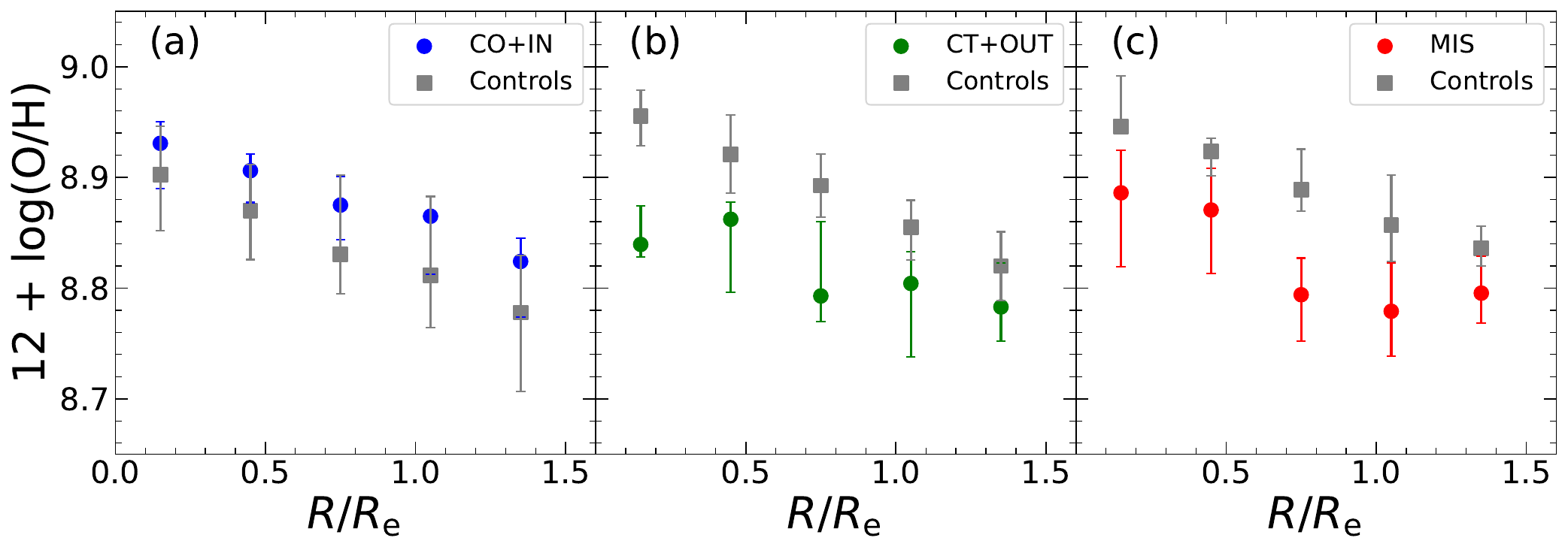}}
   \caption{The radial gradients of gas-phase metallicity. Three panels sequentially display the gas-phase metallicity radial gradients for Type~CO+IN, Type~CT+OUT, Type~MIS and their control samples. The colorful circles show the median gas-phase metallicity for different CRD types. The grey squares show the median gas-phase metallicity for their control samples. The colorful and grey error bars show 40\%-60\% scattering ranges for corresponding radial gradients.}
   \label{fig:NII_SII}
\end{figure*}

Figure~\ref{fig:NII_SII} compares the radial gradients of gas-phase metallicity between Type~CO+IN (panel a), Type~CT+OUT (panel b), Type~MIS (panel c) and their control samples, respectively. In Figure~\ref{fig:NII_SII}, the colored circles represent the median metallicity within each radial bin for CRD samples, with the error bars showing 40\% - 60\% scattering ranges. The error bar reflects the scatter in metallicity among the sample (or control) galaxies in each radial bin, with a typical value of $\sim$0.04~dex. Using the flux errors of the emission lines provided by the MaNGA DAP, we also estimated the metallicity measurement error for each spaxel via error propagation. The typical measurement error in each radial bin is $\sim$0.03~dex, which is smaller than the typical scatter in metallicity ($\sim$0.04~dex). The gray squares and error bars represent the median metallicity and scattering ranges for control samples. As shown in Figure~\ref{fig:NII_SII}(a), Type~CO+IN exhibit $\sim$0.04~dex higher gas-phase metallicity than their control sample at all radii.  In contrast, Type~CT+OUT show $\sim$0.07~dex lower gas-phase metallicity than the control sample in Figure~\ref{fig:NII_SII}(b). Meanwhile, Type~MIS in Figure~\ref{fig:NII_SII}(c) exhibit $\sim$0.07~dex lower gas-phase metallicity than their control sample. 

The observed gas-phase metallicity is the result of competition between the enrichment of stellar nucleosynthesis and the dilution of metallicity by accreting low-metallicity external gas. Considering that Type~CO+IN are dominated by SF galaxies, the $\sim$0.04~dex higher gas-phase metallicity in these galaxies compared to their controls can be attributed to enrichment from ongoing star formation. In the closed-box model \citep{2007ApJ...658..941D}, metallicity primarily depends on the gas mass fraction $f_{\mathrm{gas}} \equiv M_{\mathrm{gas}}/(M_{\mathrm{gas}} + M_\star)$; thus, abundances rise rapidly as a substantial fraction of the available gas is converted into stars. The $\sim$0.07~dex lower gas metallicity in Type~CT+OUT compared to their control sample indicates that metallicity dilution through external gas accretion is more effective than enrichment process. The misalignment between the gas disk and two stellar disks in Type~MIS suggests that the gas component is newly accreted after both stellar disks have formed. Stellar nucleosynthesis-induced metallicity enrichment can be neglected. The $\sim$0.07~dex lower gas-phase metallicity in Type~MIS compared to their control sample is primarily attributed to dilution from gas accretion.

\subsubsection{Stellar population properties}

\begin{figure*}[h]
    \centering\resizebox{0.8\textwidth}{!}{\includegraphics{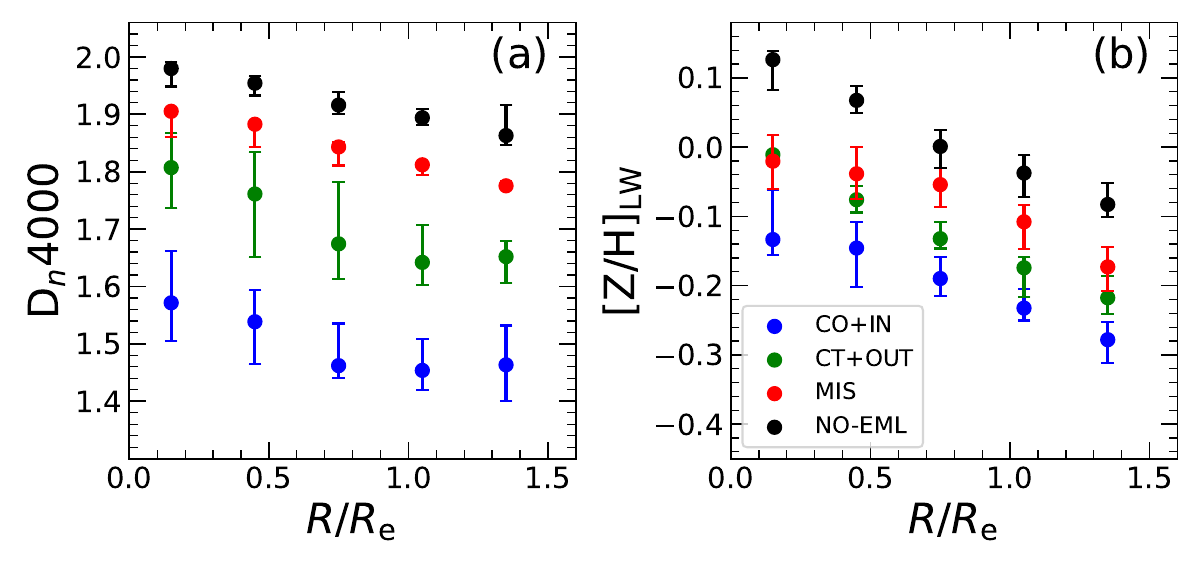}}
   \caption{The radial gradients of stellar population properties for different types of CRDs. (a) The D$_{n}$4000 radial gradients. (b) The [Z/H] radial gradients. In each panel, the blue, green, red and black circles show the median radial gradients for Type~CO+IN, Type~CT+OUT, Type~MIS and Type~NO-EML, respectively.}
   \label{fig:stellar_population}
\end{figure*}

Figure~\ref{fig:stellar_population} displays the radial gradients of D$_{n}$4000 (left) and light-weighted stellar metallicity (right). In each panel, the blue, green, red and black circles represent the median values for Type~CO+IN, Type~CT+OUT, Type~MIS and Type~NO-EML. The corresponding error bars show the 40\% - 60\% scattering ranges, following the same representation as in Figure~\ref{fig:NII_SII}. In Figure~\ref{fig:stellar_population}(a), stellar populations become older following a sequence of Type CO+IN, Type CT+OUT, Type MIS, and Type NO-EML at all radii. This trend aligns with the result in Table \ref{tbl:sequence}, where the types of CRDs with higher fractions of SF galaxies tend to host younger stellar populations. Type~CO+IN are dominated by SF galaxies with abundant pre-existing gas, the effective interaction between pre-existing and accreted gas can trigger active star formation. Meanwhile, Type~CT+OUT are dominated by QS galaxies with lower amount of pre-existing gas. This results in lower angular momentum lose efficiency through gas interaction and thus suppressing star formation activity. Type~NO-EML, which possess the highest stellar mass, oldest stellar populations, and largest fraction of QS galaxies among all the types of CRDs, may represent the final evolutionary stage of the other types. About Type~MIS, we will discuss them in Section \ref{sec:misalign} in more detail.

Figure~\ref{fig:stellar_population}(b) illustrates that the light-weighted stellar metallicity increases consistently with the sequence traced by D$_{n}$4000. By combining the trends observed in both panels, we conclude that CRDs with older stellar populations exhibit higher stellar metallicity. The higher stellar metallicity in older galaxies fundamentally arises from the process of successive generations of star formation and the metal enrichment cycle $-$ older galaxies have experienced longer duration of star formation, where each generation of stars injects heavy elements into the interstellar medium through nucleosynthesis and feedback. Subsequent star formation then utilizes these metal-enriched gas, leading to a gradual increase in stellar metallicity over time. 

\section{Discussion} \label{sec:discussion}

\subsection{Multiple gas accretion events in nearby galaxies} \label{sec:misalign}

In this section, we focus on eight CRDs classified as Type~MIS, which host two counter-rotating stellar disks and exhibit misalignment between both stellar disks and the gas disk. We propose that these galaxies underwent multiple gas accretion events with differing AM directions. So far, observations of multiply misaligned galaxies remain rare \citep{2022NatAs...6.1464C}, while these galaxies are crucial for understanding how the frequent accretion events contribute to galaxy assembly.

Globally, Type~MIS exhibit the lowest stellar mass with $\log (M_\ast/M_\odot) \sim 10.2$ and star formation rate with $\log (\mathrm{SFR}/M_\odot~yr^{-1}) \sim -4.0$  among all the types of CRDs. Additionally, they show distinct properties compared to their control sample: they have high bulge-to-total ratio with $B/T \sim 0.60$, low molecular gas mass fraction with $\log (M_{\mathrm{mol}} / M_\ast) \sim -1.5$, and reside in least dense environments with $Q_{\mathrm{LSS}} \sim -4.9$ compared to the other types of CRDs and control samples (Figure \ref{fig:Qlss}). Spatially resolved properties further reveal that Type~MIS display $\sim$0.07~dex lower gas-phase metallicity than the control sample at all radii (Figure \ref{fig:NII_SII}c) and host the second-oldest stellar populations with $\mathrm{D}_n4000 \sim 1.8$ among all the types of CRDs (Figure \ref{fig:stellar_population}a).

The significantly less dense environments of Type~MIS reduce the impacts of ram pressure stripping and halo heating, facilitating more frequent gas accretion events. Early gas accretion supplied the formation of a counter-rotating stellar disk relative to the pre-existing one. The currently observed gas disk originates from a recent accretion episode, which is independent of the early gas accretion event. The older stellar populations in Type~MIS compared to Type~CO+IN and Type~CT+OUT may be attributed to the earlier formation of their counter-rotating stellar disks. The lower molecular gas content in Type~MIS may be attributed to selection bias: early star formation efficiently depleted the gas reservoir, making these galaxies more likely to be classified as Type~MIS. This arises because the negligible interaction between recently accreted gas and pre-existing gas minimizes the impact of gas mixing, preserving the AM of the recently accreted gas. Consequently, the primary effect of this recent accretion episode is gas dilution, leading to the lower gas-phase metallicity observed in Type~MIS compared to control sample.

\cite{2022NatAs...6.1464C} identified galaxies with misaligned gas disks from the MaNGA survey, where two gas disks are also misaligned with the stellar disk. However, owing to the large collision cross-sections of gas components $-$ which result in short dynamical friction timescales ($\sim$100 million years) $-$ only two such galaxies were confirmed. In this work, eight CRDs classified as Type~MIS significantly expand the observational sample of nearby galaxies that have undergone multiple gas accretion events. Their existence demonstrates that frequent gas accretion plays a critical role in galaxy assembly, even for low-redshift galaxies, they can accrete gas multiple times from their environments. In future work, we will investigate the formation scenario for Type~MIS through numerical simulations to gain deeper theoretical insights.

\subsection{Formation scenarios for counter-rotating stellar disks}

Previous studies have established that external gas accretion is the primary formation mechanism for CRDs. The disk formed from accreted gas inherits the AM of the accreted material and counter-rotates relative to the pre-existing stellar disk. Consequently, in all previous case studies of CRDs, gas disk has been observed to co-rotate with younger stellar disk. Meanwhile, \cite{2022MNRAS.511..139B} identified two peculiar CRDs in the MaNGA survey, where gas disk co-rotates with older stellar disk. They proposed that these galaxies formed via the merger of a disk galaxy and a gas-poor galaxy hosting younger stellar populations in a retrograde orbit.

\begin{figure*}[h]
    \centering\resizebox{0.8\textwidth}{!}{\includegraphics{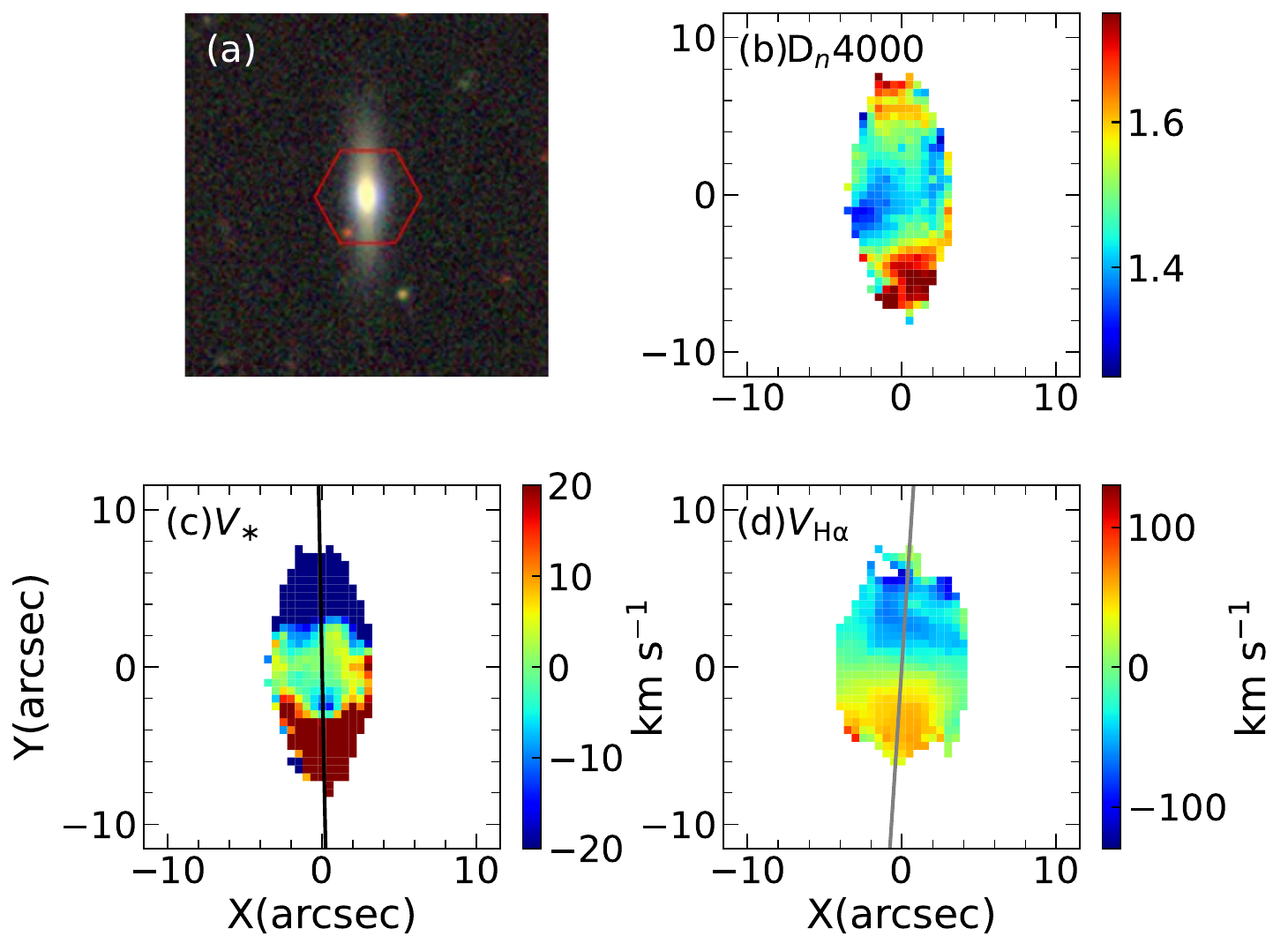}}
   \caption{An example for the peculiar CRDs, where the gas disk co-rotates with the older stellar disk. (a) DESI $g, r, z$ composite image. (b) D$_{n}$4000 map. Red color shows the relatively older stellar populations, and blue color shows the younger stellar populations. (c) Stellar velocity field. The black solid line shows the kinematic position angle of outer stellar disk. (d) The gas velocity field traced by {\ha}. The grey solid line shows the kinematic position angle of gas disk.}
   \label{fig:blue-core}
\end{figure*}

Analyzing the gas and stellar velocity fields, along with D$_{n}$4000 maps, of our CRD sample, we identify four peculiar CRDs where gas disk co-rotates with older stellar disk. Figure~\ref{fig:blue-core} presents an example of such galaxies. We further examine their DESI $g, r, z$ composite images (Figure \ref{fig:blue-core}a showing an example), which are two magnitudes deeper than SDSS observations. However, following the method of \cite{2021MNRAS.501...14L}, we detect no merger remnant or interaction feature with companion in these images.

Figures~\ref{fig:blue-core}(b), \ref{fig:blue-core}(c) and \ref{fig:blue-core}(d) present the spatially resolved D$_n$4000 map alongside the gas and stellar velocity fields for the example. Two key features are evident: (1) the inner and outer stellar disks counter-rotate, with the outer stellar disk co-rotating with the gas disk; (2) the outer stellar disk hosts older stellar populations with higher D$_n$4000. We propose an alternative formation scenario for these peculiar CRDs involving external gas accretion. Initially, the older stellar disk formed from pre-existing gas. Subsequent external gas accretion supplied counter-rotating gas to these galaxies. The interaction between the pre-existing and accreted gas redistributes the angular momentum, which further triggers gas inflows and the formation of inner stellar disk. Hence, the inner stellar disk is counter-rotating with the outer one, and hosts relatively younger stellar populations. Since the pre-existing gas has higher AM than the accreted gas, it governs the rotation of the gas disk, resulting in co-rotation between the gas disk and the outer stellar disk, which hosts relatively older stellar populations.

The stellar velocity field shown in Figure \ref{fig:blue-core}(c) reveals extremely low rotational velocities in the inner stellar disk, with absolute values below 10~km~s$^{-1}$. This suggests that our selection criteria may have overlooked CRDs featuring slowly rotating inner stellar disk, indicating that the 1.5\% CRD detection rate should be considered as a lower limit for the incidence rate for CRDs in the MaNGA survey. \cite{2014ApJ...792L...4P} investigated nearby green-valley galaxies, and detected blue-core in 63\% of the early-type green-valley galaxies, suggesting that gas inflows can fuel star formation in their center. \cite{2019RAA....19...81C} constructed a sample of blue-core galaxies from the MaNGA survey, and found that 26\% of these galaxies exhibit misalignment between gas and stellar components, indicating external gas accretion. These blue-core galaxies with younger stellar populations in the center may have undergone a formation scenario similar to the CRDs in Figure \ref{fig:blue-core}. Their exclusion from the CRD sample can be attributed to the exceptionally low rotational velocities in the inner stellar disk.

\section{Conclusion} \label{sec:conclusion}

In this study, we construct a sample of 147 galaxies with counter-rotating stellar disks from the MaNGA survey, and 138 out of them have global $M_\ast$ and SFR measurements from the catalog of \cite{2015ApJS..219....8C}. We classify these 138 CRDs into six distinct types, compare global and spatially resolved properties between CRDs and their control samples with similar stellar mass and star formation rate. The main results are as follows:

\begin{itemize}

    \item[1.] All the types of CRDs exhibit more prominently bulge-dominated morphologies, lower molecular gas mass fraction, and reside in less dense environments compared to their control samples. These features support a mechanism that the formation of CRDs is dominated by external gas accretion.

    \item[2.] Type~CO+IN are domimated by star-forming galaxies with abundant pre-existing gas in their progenitors. The interaction between pre-existing and accreted gas triggers active star formation, resulting in the youngest stellar populations among all the types of CRDs. In contrast, Type~CT+OUT are dominated by quiescent galaxies, hosting older stellar populations than Type~CO+IN. Among all the types of CRDs, Type~NO-EML have the oldest stellar populations and highest stellar mass, suggesting that they may represent the final evolutionary stage of the other types.

    \item[3.] For Type~MIS, the fact that both stellar disks are misaligned with the gas disk implies that their formation is governed by multiple gas accretion events with differing AM direction of the accreted gas. We propose that early gas accretion triggered the formation of a counter-rotating stellar disk relative to the pre-existing one, while the recently accreted gas governs the rotation of the gas disk.

    \item[4.] We identify four CRDs where the gas disk co-rotates with the older stellar disk. \cite{2022MNRAS.511..139B} suggested galaxy merger as a formation mechanism. However, we find no evidence of merger remnant or interaction feature with companion in the DESI images of these galaxies. In addition to the galaxy merger, we propose another possible mechanism through external gas accretion, where the pre-existing gas has higher AM than the accreted gas, therefore governing the gas disk's rotation.

\end{itemize}

\appendix

\begin{longtable}{c c c c c c c c c} 
    \caption{Galaxies with counter-rotating stellar disks in MaNGA survey.}\\       
    \hline
    \hline
    MaNGA-ID & Plate & IFU & R.A. & Dec. & z & $\log M_\ast$ (M$_\odot$) & $\log \mathrm{SFR}$ ($\mathrm{M}_\odot~\mathrm{yr}^{-1}$) & Class \\
    \text{(1)} & \text{(2)} & \text{(3)} & \text{(4)} & \text{(5)} & \text{(6)} & \text{(7)} & \text{(8)} & \text{(9)} \\
    \hline
    1-109056 & 8077 & 6103 & 39.447 & 0.405 & 0.047 & 10.600 & -1.988 & Type CO \\
    1-39396 & 8091 & 6102 & 12.291 & 14.644 & 0.040 & 9.970 & -0.298 & Type CO \\
    1-44183 & 8138 & 3704 & 118.020 & 44.161 & 0.026 & 9.930 & -0.643 & Type CO \\
    1-37494 & 8154 & 3703 & 44.625 & 0.794 & 0.043 & 10.260 & -3.533 & Type CO \\
    1-37155 & 8154 & 6103 & 44.349 & -0.309 & 0.029 & 11.010 & -2.643 & Type CO \\
    1-52701 & 8158 & 6102 & 61.406 & -5.315 & 0.033 & 10.750 & -3.363 & Type CO \\
    1-72795 & 8240 & 3701 & 124.736 & 44.320 & 0.117 & 11.410 & -0.108 & Type CO \\
    1-277289 & 8255 & 3704 & 165.275 & 45.293 & 0.037 & 9.860 & -1.388 & Type CO \\
    1-516580 & 8310 & 1902 & 180.255 & 23.725 & 0.030 & 9.610 & -0.253 & Type CO \\
    1-259068 & 8341 & 6102 & 190.353 & 45.046 & 0.025 & 10.060 & 0.232 & Type CO \\
    1-260458 & 8447 & 9101 & 206.070 & 39.509 & 0.062 & 11.090 & -3.883 & Type CO \\
    1-260823 & 8447 & 3703 & 208.292 & 40.119 & 0.051 & 10.640 & -2.078 & Type CO \\
    1-274545 & 8455 & 3704 & 157.686 & 40.057 & 0.023 & 9.590 & -0.213 & Type CO \\
    1-134361 & 8487 & 1902 & 240.224 & 47.197 & 0.046 & 10.520 & -4.398 & Type CO \\
    1-210962 & 8550 & 12701 & 246.359 & 39.871 & 0.029 & 11.150 & -1.338 & Type CO \\
    1-134737 & 8600 & 3701 & 245.912 & 41.834 & 0.027 & 9.580 & -1.668 & Type CO \\
    1-25632 & 8611 & 9101 & 260.674 & 59.852 & 0.027 & 10.600 & -4.203 & Type CO \\
    1-546819 & 8614 & 3703 & 257.001 & 36.344 & 0.036 & 11.130 & -3.753 & Type CO \\
    1-24279 & 8626 & 6102 & 263.173 & 56.117 & 0.063 & 11.330 & -3.583 & Type CO \\
    1-44312 & 8718 & 1902 & 119.638 & 45.329 & 0.022 & 9.330 & -1.093 & Type CO \\
    1-45520 & 8725 & 3701 & 125.274 & 45.848 & 0.038 & 9.580 & -0.028 & Type CO \\
    1-457547 & 8982 & 1902 & 201.921 & 26.062 & 0.024 & 9.970 & -1.803 & Type CO \\
    1-148987 & 8999 & 6102 & 164.552 & 50.213 & 0.024 & 10.410 & -4.438 & Type CO \\
    1-323886 & 9027 & 3702 & 245.142 & 32.935 & 0.041 & 10.040 & -0.278 & Type CO \\
    1-323764 & 9027 & 3703 & 246.054 & 31.876 & 0.021 & 9.550 & -0.823 & Type CO \\
    1-42660 & 9195 & 3704 & 28.185 & 13.193 & 0.026 & 10.240 & -0.443 & Type CO \\
    1-121837 & 9485 & 3702 & 120.076 & 37.609 & 0.042 & 10.210 & -1.638 & Type CO \\
    1-121871 & 9485 & 3703 & 120.767 & 36.264 & 0.044 & 10.140 & -0.503 & Type CO \\
    1-386322 & 9506 & 3703 & 134.214 & 27.270 & 0.019 & 10.030 & -4.933 & Type CO \\
    1-537081 & 9882 & 3702 & 206.949 & 23.367 & 0.084 & 11.050 & 0.337 & Type CO \\
    1-317315 & 9888 & 1902 & 238.028 & 27.559 & 0.072 & 11.140 & -3.793 & Type CO \\
    1-456645 & 11009 & 12703 & 195.775 & 26.531 & 0.019 & 9.930 & -0.573 & Type CO \\
    1-245176 & 11013 & 1902 & 209.898 & 55.593 & 0.041 & 10.180 & -0.133 & Type CO \\
    1-78294 & 11753 & 6104 & 147.068 & 3.906 & 0.024 & 10.120 & 0.172 & Type CO \\
    1-407465 & 11826 & 6104 & 190.001 & 36.872 & 0.065 & 10.810 & 0.317 & Type CO \\
    1-234243 & 11830 & 3702 & 197.696 & 50.093 & 0.031 & 10.920 & 0.057 & Type CO \\
    1-1243 & 11836 & 1902 & 150.717 & 0.069 & 0.032 & 10.080 & -2.378 & Type CO \\
    \hline
    1-135767 & 11946 & 12703 & 250.229 & 37.904 & 0.031 & 10.560 & -4.138 & Type CO \\
    1-139814 & 12489 & 3703 & 179.348 & 4.543 & 0.020 & 10.040 & -0.443 & Type CO \\
    1-2880 & 12518 & 3704 & 159.894 & -0.253 & 0.019 & 9.160 & -0.563 & Type CO \\
    1-97176 & 12679 & 3701 & 318.291 & -7.632 & 0.028 & 10.510 & -4.458 & Type CO \\
    1-145913 & 8149 & 1901 & 120.222 & 28.130 & 0.017 & 9.900 & -4.998 & Type IN \\
    1-137890 & 8249 & 1901 & 137.219 & 44.932 & 0.027 & 10.060 & 0.037 & Type IN \\
    1-255220 & 8253 & 1902 & 157.846 & 42.277 & 0.022 & 9.920 & -4.998 & Type IN \\
    1-274440 & 8455 & 1902 & 155.709 & 39.369 & 0.026 & 9.540 & -0.533 & Type IN \\
    1-634718 & 8610 & 3704 & 259.630 & 60.586 & 0.013 & 9.530 & -2.408 & Type IN \\
    1-179561 & 8615 & 1902 & 319.751 & -0.964 & 0.020 & 10.020 & -0.113 & Type IN \\
    1-26197 & 8652 & 6102 & 330.724 & -0.785 & 0.064 & 11.130 & -3.828 & Type IN \\
    1-188530 & 8995 & 3703 & 175.818 & 55.278 & 0.055 & 10.780 & -4.158 & Type IN \\
    1-323766 & 9027 & 1902 & 246.176 & 32.064 & 0.022 & 9.900 & -2.393 & Type IN \\
    1-295839 & 9047 & 3703 & 247.072 & 25.704 & 0.131 & 11.430 & -3.293 & Type IN \\
    1-109275 & 9192 & 3703 & 46.071 & 0.520 & 0.044 & 10.060 & -0.308 & Type IN \\
    1-42247 & 9514 & 1902 & 33.588 & 14.119 & 0.029 & 9.710 & -0.143 & Type IN \\
    1-42082 & 9514 & 3701 & 32.825 & 12.922 & 0.060 & 10.880 & -4.073 & Type IN \\
    1-593328 & 9868 & 12704 & 219.657 & 46.663 & 0.037 & nan & nan & Type IN \\
    1-633000 & 9872 & 3701 & 233.232 & 42.438 & 0.020 & 10.110 & -0.103 & Type IN \\
    1-389466 & 10838 & 1902 & 147.550 & 34.699 & 0.038 & 10.090 & -0.028 & Type IN \\
    1-245235 & 11012 & 3702 & 211.425 & 55.777 & 0.041 & 10.150 & 0.022 & Type IN \\
    1-13736 & 11834 & 1902 & 223.081 & -0.269 & 0.044 & 10.080 & -0.453 & Type IN \\
    1-22347 & 7992 & 3701 & 252.171 & 63.479 & 0.044 & 10.430 & -4.513 & Type CT \\
    1-339061 & 8138 & 6102 & 117.144 & 44.688 & 0.020 & 10.410 & -4.093 & Type CT \\
    1-38543 & 8155 & 3702 & 54.032 & -0.596 & 0.023 & 10.500 & -4.428 & Type CT \\
    1-199775 & 8314 & 3702 & 240.848 & 39.986 & 0.030 & 10.170 & -0.063 & Type CT \\
    1-284335 & 8318 & 6103 & 196.821 & 45.728 & 0.035 & 9.840 & -4.873 & Type CT \\
    1-251198 & 8332 & 3704 & 209.640 & 41.724 & 0.043 & 10.220 & -0.453 & Type CT \\
    1-419257 & 8444 & 3701 & 201.141 & 31.546 & 0.023 & 10.260 & -0.823 & Type CT \\
    1-418253 & 8446 & 1902 & 206.154 & 37.172 & 0.027 & 10.650 & -4.278 & Type CT \\
    1-210611 & 8600 & 1902 & 244.391 & 41.689 & 0.027 & nan & nan & Type CT \\
    1-379008 & 8711 & 1902 & 118.761 & 52.227 & 0.023 & 9.710 & -1.063 & Type CT \\
    1-71956 & 8713 & 3704 & 118.442 & 38.869 & 0.040 & 9.830 & -2.323 & Type CT \\
    1-188177 & 8991 & 6103 & 176.130 & 53.683 & 0.027 & 10.430 & -4.498 & Type CT \\
    1-149871 & 8992 & 3704 & 173.487 & 51.250 & 0.026 & 10.840 & -4.108 & Type CT \\
    1-94690 & 9026 & 3704 & 251.378 & 43.582 & 0.031 & 10.310 & -0.853 & Type CT \\
    1-94773 & 9026 & 1902 & 249.302 & 44.086 & 0.033 & 10.180 & -1.448 & Type CT \\
    1-295542 & 9048 & 1902 & 246.256 & 24.263 & 0.050 & 10.450 & 0.737 & Type CT \\
    1-178027 & 9186 & 3704 & 258.743 & 29.138 & 0.045 & 10.540 & -4.333 & Type CT \\
    1-382889 & 9496 & 3701 & 120.620 & 20.514 & 0.029 & 10.580 & -1.043 & Type CT \\
    1-153127 & 10220 & 1901 & 120.805 & 33.132 & 0.018 & 9.880 & -4.903 & Type CT \\
    1-76425 & 10511 & 3704 & 131.828 & 2.762 & 0.028 & 10.760 & -4.153 & Type CT \\
    1-77006 & 10512 & 1902 & 135.588 & 3.616 & 0.024 & 10.030 & -1.748 & Type CT \\
    1-151874 & 11019 & 1901 & 192.363 & 51.582 & 0.032 & 10.380 & -2.628 & Type CT \\
    1-244115 & 11020 & 1901 & 204.506 & 55.188 & 0.025 & 10.380 & -3.183 & Type CT \\
    1-586725 & 11838 & 3704 & 156.032 & 0.142 & 0.023 & nan & nan & Type CT \\
    1-135190 & 11942 & 3702 & 246.753 & 40.811 & 0.030 & 10.630 & -4.298 & Type CT \\
    1-294016 & 11981 & 1901 & 254.321 & 19.339 & 0.033 & 9.560 & -0.628 & Type CT \\
    1-386932 & 12487 & 1902 & 137.844 & 30.096 & 0.026 & 9.740 & -0.413 & Type CT \\
    \hline
    1-41258 & 8093 & 3703 & 22.175 & 14.609 & 0.036 & 10.870 & -4.073 & Type OUT \\
    1-44483 & 8143 & 3702 & 119.835 & 42.057 & 0.025 & 10.320 & -4.618 & Type OUT \\
    1-44047 & 8143 & 1902 & 119.643 & 41.578 & 0.041 & 10.260 & -3.843 & Type OUT \\
    1-232087 & 8152 & 3703 & 142.725 & 35.440 & 0.028 & 10.390 & -4.563 & Type OUT \\
    1-251067 & 8332 & 3703 & 207.478 & 43.482 & 0.045 & 10.380 & -4.533 & Type OUT \\
    1-266244 & 8333 & 1902 & 216.547 & 42.711 & 0.017 & 9.010 & -0.743 & Type OUT \\
    1-136248 & 8606 & 3702 & 253.794 & 36.906 & 0.024 & 10.580 & -3.688 & Type OUT \\
    1-248410 & 8979 & 3701 & 241.209 & 42.036 & 0.025 & 9.660 & -0.753 & Type OUT \\
    1-174947 & 8989 & 9101 & 176.359 & 49.879 & 0.033 & 10.700 & -2.998 & Type OUT \\
    1-269003 & 9032 & 1902 & 242.045 & 31.515 & 0.014 & 9.750 & -4.998 & Type OUT \\
    1-78381 & 10515 & 12702 & 148.003 & 4.151 & 0.031 & 11.090 & -3.848 & Type OUT \\
    1-245301 & 11012 & 6104 & 212.532 & 54.920 & 0.040 & 10.880 & -4.038 & Type OUT \\
    1-566335 & 11962 & 3701 & 226.283 & 8.107 & 0.047 & nan & nan & Type OUT \\
    1-40771 & 12093 & 3701 & 17.785 & 14.223 & 0.054 & 11.280 & -3.653 & Type OUT \\
    1-113520 & 7815 & 1901 & 317.502 & 11.511 & 0.017 & 9.970 & -4.998 & Type MIS \\
    1-519907 & 8310 & 1901 & 179.645 & 24.117 & 0.030 & 10.130 & -4.198 & Type MIS \\
    1-418023 & 8446 & 1901 & 205.753 & 36.166 & 0.024 & nan & nan & Type MIS \\
    1-248869 & 8604 & 6103 & 244.691 & 39.334 & 0.032 & 11.020 & -3.923 & Type MIS \\
    1-635506 & 8618 & 6102 & 319.103 & 10.273 & 0.017 & 10.200 & -1.618 & Type MIS \\
    1-163594 & 8726 & 1901 & 115.494 & 23.176 & 0.043 & 11.140 & -2.493 & Type MIS \\
    1-235983 & 8980 & 3701 & 223.636 & 42.385 & 0.018 & 9.860 & -4.998 & Type MIS \\
    1-246484 & 9865 & 1902 & 223.866 & 51.048 & 0.030 & 9.910 & -4.968 & Type MIS \\
    1-4109 & 12491 & 1901 & 166.240 & 1.038 & 0.043 & 10.130 & -0.653 & Type MIS \\
    1-145871 & 8148 & 3702 & 118.531 & 27.692 & 0.065 & 11.060 & -3.903 & Type NO-EML \\
    1-37478 & 8154 & 1902 & 44.847 & -1.167 & 0.028 & 10.560 & -3.998 & Type NO-EML \\
    1-196580 & 8309 & 6102 & 208.508 & 51.905 & 0.144 & 11.550 & -3.418 & Type NO-EML \\
    1-250784 & 8327 & 1902 & 202.961 & 44.152 & 0.034 & 10.910 & -2.708 & Type NO-EML \\
    1-251783 & 8335 & 1901 & 215.311 & 39.653 & 0.026 & 10.240 & -1.993 & Type NO-EML \\
    1-166613 & 8461 & 3701 & 144.516 & 42.974 & 0.047 & 11.140 & -2.473 & Type NO-EML \\
    1-90192 & 8553 & 6102 & 233.713 & 56.451 & 0.039 & 11.220 & -1.153 & Type NO-EML \\
    1-95120 & 8601 & 9102 & 250.406 & 40.163 & 0.034 & 11.250 & -3.618 & Type NO-EML \\
    1-210728 & 8604 & 3701 & 247.280 & 39.496 & 0.029 & 10.490 & -4.448 & Type NO-EML \\
    1-635590 & 8615 & 3702 & 321.054 & 1.118 & 0.049 & nan & nan & Type NO-EML \\
    1-44113 & 8718 & 3701 & 120.206 & 44.691 & 0.102 & 11.610 & -3.358 & Type NO-EML \\
    1-456904 & 8933 & 3703 & 195.184 & 28.337 & 0.026 & 10.570 & -2.743 & Type NO-EML \\
    1-279554 & 8945 & 3701 & 171.628 & 46.998 & 0.025 & 10.530 & -4.298 & Type NO-EML \\
    1-624035 & 8949 & 3703 & 195.875 & 28.275 & 0.021 & 9.560 & -4.998 & Type NO-EML \\
    1-278079 & 8993 & 3704 & 166.087 & 46.056 & 0.143 & 11.510 & -3.403 & Type NO-EML \\
    1-188165 & 8995 & 3704 & 175.602 & 54.774 & 0.058 & 11.400 & -3.558 & Type NO-EML \\
    1-269227 & 9028 & 1901 & 242.670 & 30.255 & 0.031 & 10.570 & -4.368 & Type NO-EML \\
    1-135244 & 9029 & 3703 & 246.907 & 42.638 & 0.031 & 10.880 & -3.228 & Type NO-EML \\
    1-593972 & 9042 & 3703 & 235.202 & 28.292 & 0.033 & nan & nan & Type NO-EML \\
    1-50956 & 9188 & 3702 & 48.523 & -8.457 & 0.034 & 10.940 & -3.798 & Type NO-EML \\
    1-603965 & 9194 & 3701 & 46.885 & -0.965 & 0.039 & 11.380 & -1.028 & Type NO-EML \\
    1-387045 & 9505 & 1902 & 139.777 & 27.915 & 0.028 & 9.810 & -2.578 & Type NO-EML \\
    1-623416 & 9873 & 12703 & 194.475 & 28.500 & 0.024 & 10.240 & -4.293 & Type NO-EML \\
    1-456672 & 9874 & 12705 & 195.080 & 27.554 & 0.020 & 10.260 & -4.678 & Type NO-EML \\
    1-456884 & 9875 & 9101 & 194.580 & 27.762 & 0.018 & 9.570 & -4.538 & Type NO-EML \\
    1-316773 & 9888 & 3702 & 235.375 & 27.987 & 0.032 & 11.070 & -3.823 & Type NO-EML \\
    \hline
    1-297863 & 10216 & 1902 & 116.767 & 18.138 & 0.045 & 10.280 & -1.793 & Type NO-EML \\
    1-280716 & 10507 & 3703 & 177.397 & 47.831 & 0.025 & 10.480 & -4.263 & Type NO-EML \\
    1-182730 & 10519 & 3704 & 154.598 & 5.736 & 0.074 & 11.120 & -3.758 & Type NO-EML \\
    1-762 & 10843 & 3704 & 149.655 & -0.930 & 0.061 & 11.420 & -3.513 & Type NO-EML \\
    1-320459 & 11016 & 6102 & 210.887 & 49.969 & 0.071 & 11.450 & -3.533 & Type NO-EML \\
    1-559761 & 11022 & 3701 & 218.935 & 4.993 & 0.030 & nan & nan & Type NO-EML \\
    1-83002 & 11835 & 9102 & 221.318 & 3.404 & 0.028 & 10.610 & -3.973 & Type NO-EML \\
    1-283130 & 11953 & 3701 & 189.060 & 42.637 & 0.028 & 9.800 & -2.003 & Type NO-EML \\
    1-333770 & 11962 & 6103 & 226.307 & 8.797 & 0.045 & 11.110 & -3.058 & Type NO-EML \\
    1-632581 & 11967 & 3701 & 230.913 & 8.759 & 0.034 & 10.920 & -3.993 & Type NO-EML \\
    1-591742 & 12620 & 1901 & 201.115 & 30.978 & 0.023 & nan & nan & Type NO-EML \\
    1-135772 & 12667 & 3703 & 250.386 & 37.266 & 0.101 & 11.420 & -3.528 & Type NO-EML \\
    \hline
    \label{tbl:sample}
\end{longtable}
\vspace{-20pt}
{\small (1) MaNGA-ID; (2) Plate; (3) IFU; (4) Right Ascension; (5) Declination; (6) Redshift; (7) Stellar mass; (8) Star formation rate; (9) Classification of the CRD galaxies based on their stellar and gas kinematics.}

\vspace{20pt}

\textbf{Acknowledgements:} M.B. acknowledges support by the National Natural Science Foundation of China, NSFC grants No. 12541302 and No. 12303009. Y.M.C. acknowledges support by NSFC grant No. 12333002, the China Manned Space Project No. CMS-CSST-2025-A08 and the Fundamental Research Funds for the Central Universities No. KG202502. Funding for the Sloan Digital Sky Survey IV has been provided by the Alfred P. Sloan Foundation, the U.S. Department of Energy Office of Science, and the Participating Institutions. SDSS- IV acknowledges support and resources from the Center for High-Performance Computing at the University of Utah. The SDSS web site is www.sdss.org. SDSS-IV is managed by the Astrophysical Research Consortium for the Participating Institutions of the SDSS Collaboration including the Brazilian Participation Group, the Carnegie Institution for Science, Carnegie Mellon University, the Chilean Participation Group, the French Participation Group, Harvard-Smithsonian Center for Astrophysics, Instituto de Astrof\'{i}sica de Canarias, The Johns Hopkins University, Kavli Institute for the Physics and Mathematics of the Universe (IPMU) / University of Tokyo, Lawrence Berkeley National Laboratory, Leibniz Institut  f\"{u}r Astrophysik Potsdam (AIP), Max-Planck-Institut  f\"{u}r   Astronomie  (MPIA  Heidelberg), Max-Planck-Institut   f\"{u}r   Astrophysik  (MPA   Garching), Max-Planck-Institut f\"{u}r Extraterrestrische Physik (MPE), National Astronomical Observatory of China, New Mexico State University, New York University, University of Notre Dame, Observat\'{o}rio Nacional / MCTI, The Ohio State University, Pennsylvania State University, Shanghai Astronomical Observatory, United Kingdom Participation Group, Universidad Nacional  Aut\'{o}noma de M\'{e}xico,  University of Arizona, University of Colorado  Boulder, University of Oxford, University of Portsmouth, University of Utah, University of Virginia, University  of Washington,  University of  Wisconsin, Vanderbilt University, and Yale University.

\end{document}